\documentclass[aps, prl, twocolumn, a4paper, floatfix]{revtex4-1}

\usepackage[utf8]{inputenc}
\usepackage{epsfig}
\usepackage[T1]{fontenc}
\usepackage{t1enc}
\usepackage{graphicx}
\usepackage{amssymb}
\usepackage{amsmath}

\usepackage{relsize}
\newcommand{\avg}[1]{{\left<#1\right>}}

\usepackage{mathtools}
\def\multiset#1#2{\ensuremath{\left(\kern-.3em\left(\genfrac{}{}{0pt}{}{#1}{#2}\right)\kern-.3em\right)}}
\newcommand{\drawbox}[1]{#1}

\usepackage{pdfpages}

\begin{document}

\title{Evolution of robust network topologies: Emergence of central backbones}

\author{Tiago P. Peixoto}
\email{tiago@itp.uni-bremen.de}
\affiliation{Institut f\"{u}r Theoretische Physik, Universit\"at Bremen, Hochschulring 18, D-28359 Bremen, Germany}
\author{Stefan Bornholdt}
\email{bornholdt@itp.uni-bremen.de}
\affiliation{Institut f\"{u}r Theoretische Physik, Universit\"at Bremen, Hochschulring 18, D-28359 Bremen, Germany}

\pacs{05.40.-a, 05.40.Ca, 05.70.Fh, 02.50.Cw, 02.30.Sa, 87.16.Yc, 87.18.Cf, 89.75.-k, 89.75.Hc }

\begin{abstract}
  We model the robustness against random failure or intentional attack
  of networks with arbitrary large-scale structure. We construct a
  block-based model which incorporates --- in a general fashion --- both
  connectivity and interdependence links, as well as arbitrary degree
  distributions and block correlations. By optimizing the percolation
  properties of this general class of networks, we identify a simple
  core-periphery structure as the topology most robust against random
  failure. In such networks, a distinct and small ``core'' of nodes with
  higher degree is responsible for most of the connectivity, functioning
  as a central ``backbone'' of the system. This centralized topology
  remains the optimal structure when other constraints are imposed, such
  as a given fraction of interdependence links and fixed degree
  distributions.  This distinguishes simple centralized topologies as
  the most likely to emerge, when robustness against failure is the
  dominant evolutionary force.
\end{abstract}

\maketitle

One of the most important characteristics of large networked systems is
their capacity to function reliably. Perhaps the simplest paradigm
employed in the study of robustness of network systems is the theory of
\emph{percolation}~\cite{callaway_network_2000}, which describes the
conditions for the formation of a system-spanning connected
component. Arguably, if a system is not connected to begin with, it is
unlikely to function properly, regardless of its intended
purpose. Recently~\cite{buldyrev_catastrophic_2010}, it has been shown
that if the \emph{interdependence} of the many elements is taken into
account, the random failure of a fraction of the system can cause a
catastrophic outcome, where the relative size of the connected cluster
falls discontinuously to zero, which otherwise would have been a
continuous transition, in the absence of interdependence.

In this Letter, we investigate analytically and numerically the most
fundamental large-scale structures~\cite{newman_communities_2012} which
results in robustness against random failure and malicious attack. By
constructing a general block-based model, and analytically deriving its
percolation properties, we obtain the topological configuration which
optimizes a well-defined robustness criterion. In the case of random
failure, we find that a remarkably simple core-periphery (CP)
topology~\cite{holme_core-periphery_2005, borgatti_models_2000} emerges
as the most optimal, when no other constraints are imposed other than
the overall cost in realizing the system. This topology remains optimal
when one introduces interdependence links, and it can entirely suppress
the catastrophic failure present in fully random topologies. However, in
the case of targeted attacks, the fully random configuration turns out
to be the most robust. We also consider the scenario where degree
constraints are present, and find that the bimodal core-periphery
structure is replaced by strongly dissortative and assortative block
topologies, for random and targeted attacks, respectively.

Here we consider a \emph{stochastic
blockmodel}~\cite{holland_stochastic_1983, karrer_stochastic_2011},
which defines an ensemble of networks composed of $B$ discrete node
blocks, where $n_r$ is the number of nodes in block $r$ and $e_{rs}$ is
the number of edges between blocks $r$ and $s$ (or twice that number if
$r=s$). Each block is allowed its own degree distribution, $p^r_k$.  We
also consider interdependence edges by defining a distinct matrix
$\hat{e}_{rs}$ and degree distributions $p^r_{\hat{k}}$, in an analogous
fashion. As in~\cite{buldyrev_catastrophic_2010}, if a node $u$ is
dependent of another node $v$, $u$ fails automatically when $v$ fails,
and vice-versa. In the case of multiple
dependencies~\cite{shao_cascade_2011}, \emph{all} support nodes must
fail in order for the dependent node to fail (i.e. they are
redundant). Assuming that the values of $n_r$ are large enough, this
ensemble becomes locally tree-like~\footnote{By using a locally
tree-like ensemble from the beginning, we are \emph{a priori} ruling out
optimized structures which may not possess this property. While we cannot
in principle strictly rule out optimal non-tree-like structures, it is
known that the effect of increased clustering is in general to
\emph{reduce} the size of the percolating
component~\cite{newman_random_2009}. We therefore assume this not to be
an issue, which in fact confirmed by empirical optimizations without
this imposition~\cite{schneider_mitigation_2011}, which do not result in
clustered networks.}, thus it is possible to adapt the epidemics-based
generating function formalism~\cite{newman_random_2001,
parshani_interdependent_2010, son_percolation_2012}, which becomes exact
in the limit of large networks. If we define $u_r$ as the probability
that a node belonging to block $r$ is not in a macroscopic component via
one of its neighbors, the dilution variable $\phi^r_k$ as the fraction
of nodes belonging to block $r$ and with degree $k$ which are not
removed from the network, and additionally the \emph{interdependence
dilution} $\hat{\phi}_r$ defined as the fraction of nodes from block $r$
which were not removed due to the failure of the support nodes, we can
write the following self-consistency equations (see \emph{Supplemental
Material} for a derivation),
\begin{align}
  u_r &= \sum_s m_{rs} \left[1 - \hat{\phi}_sf_1^s(1) + \hat{\phi}_sf_1^s(u_s)\right] \label{eq:mf_u}\\
  \hat{\phi}_r &= 1 - \hat{f}^r_0(1 - {\textstyle\sum_s}\hat{m}_{rs}S^0_s) + \hat{f}^r_0(0), \label{eq:mf_phi}
\end{align}
where $m_{rs} = e_{rs} / n_r\kappa_r$ is the asymmetric matrix defining
the fraction of edges adjacent to vertices of block $r$ which are also
adjacent to block $s$, where $ \kappa_r = \sum_{s}e_{rs} / n_r$ is the
average degree of block $r$. $S_r = \hat{\phi}_rS^0_r$ is the fraction
of nodes of block $r$ which belong to a macroscopic component, and
$S^0_r = f_0^r(1) - f_0^r(u_r)$, given as a function of the diluted
degree generating function $f_0^r(z)=\sum_kp_k^r\phi^r_kz^k$.
Furthermore, $f^r_1(z) = \sum_kq_k\phi^r_{k+1} z^k$ generates the
diluted excess degree distribution of block $r$, with $q^r_k =
p^r_{k+1}(k + 1)/\kappa_r$. Note that we have $f^r_1(z) = {f^r}'_0(z) /
{g^r}'_0(1)$ where $g_0^r=\sum_kp_k^rz^k$ generates the undiluted degree
distribution of block $r$. The generating function
$\hat{f}_0^r(z)=\sum_{\hat{k}}p_{\hat{k}}^rz^{\hat{k}}$ describes the
degrees corresponding to interdependent edges alone. The total fraction
of nodes which belong to a macroscopic component is simply $S =
\sum_rw_rS_r$, where $w_r=n_r/N$ is the fraction of nodes belonging to
block $r$, and the total dilution in the network is likewise $\phi =
\sum_{r,k}w_rp^r_k\phi^r_k$~\footnote{One can modify slightly
Eqs.~\ref{eq:mf_u} and~\ref{eq:mf_phi} to cover the case of unidirected
and/or non-redundant interdependence as well, see \emph{Supplemental
Material}.}.

\begin{figure}
  \begin{minipage}[b]{0.325\columnwidth}
    \vspace{0pt}
    (a)\includegraphics[width=0.9\columnwidth]{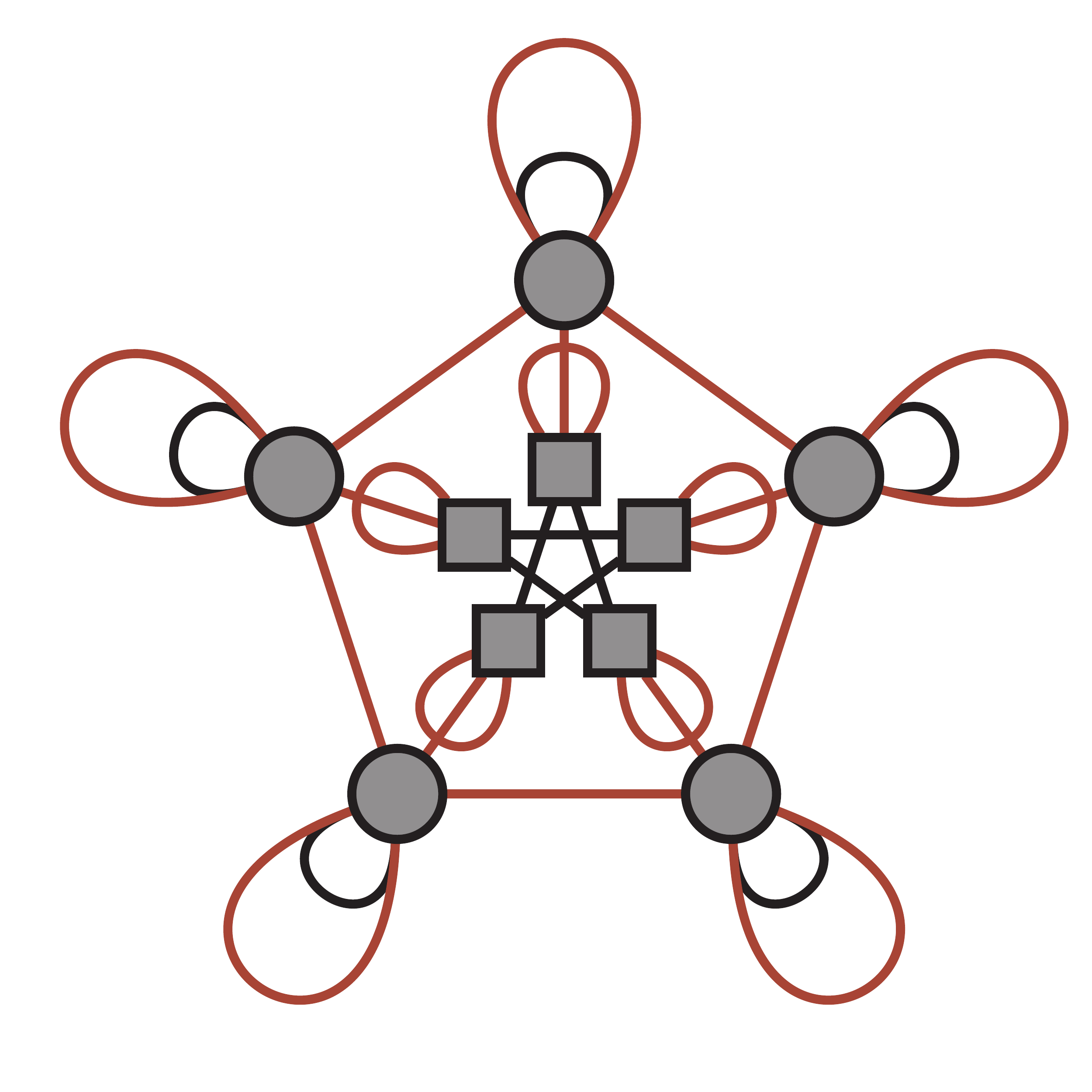}
  \end{minipage}
  \begin{minipage}[b]{0.325\columnwidth}
    \vspace{0pt}
    (b)\includegraphics[width=0.9\columnwidth]{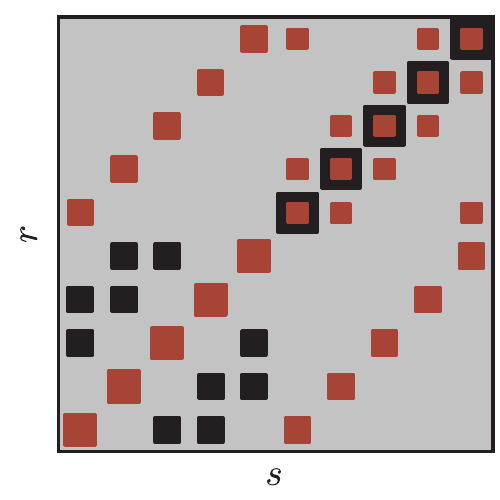}
  \end{minipage}
  \begin{minipage}[b]{0.325\columnwidth}
    \vspace{0pt}
    (c)\includegraphics[width=0.9\columnwidth,trim=0.cm 0.65cm 0.cm 0.cm]{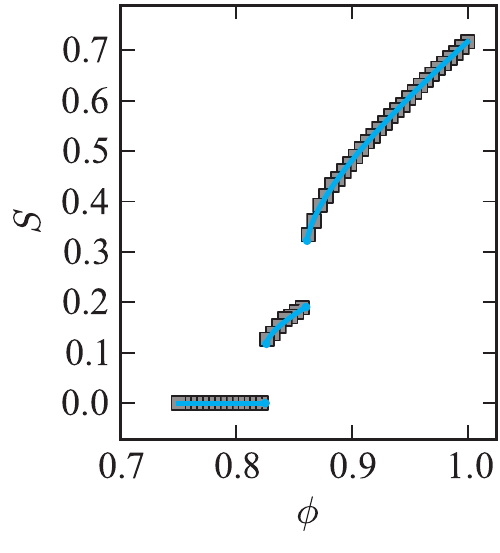}
  \end{minipage} \caption{\label{fig:example}(a) Block structure with 10
  blocks of equal size. Each node represents a block, and black (red)
  edges correspond to connectivity (interdependence) edges. Squares
  (circles) have exponentially (Poisson) distributed connectivity
  degrees with $\kappa_i = 2.25$ ($\kappa_i = 1.75$), and
  $\hat{\kappa}_i = 8$ ($\hat{\kappa}_i = 6$). All interdependence
  degrees are exponentially distributed. (b) The matrices $e_{rs}$
  (black squares) and $\hat{e}_{rs}$ (red squares) that correspond to
  this ensemble. (c) $S$ a function of $\phi$, for random failures. The
  solid lines are results obtained via Eqs.~\ref{eq:mf_u}
  and~\ref{eq:mf_phi}, and the filled symbols are obtained empirically
  with a network realization of size $N=10^6$. The two discontinuous
  jumps, correspond to the outer and inner rings failing in sequence.}

\end{figure}

Although quite straightforward, this parametrization is a generalization
of many scenarios considered in the literature, such as two fully
interdependent networks~\cite{buldyrev_catastrophic_2010}, the ``network
of networks'' topology~\cite{gao_robustness_2011, gao_robustness_2011-1,
gao_networks_2012}, single networks with mixed connectivity and
interdependence edges~\cite{bashan_percolation_2011, hu_percolation_2011,
parshani_critical_2011}, as well as networks with only connectivity
edges and arbitrary degree distribution and
correlations~\cite{vazquez_resilience_2003}, all of which correspond to
specific choices of the parameters defined here. In
Fig.~\ref{fig:example} is shown an example of a block structure with
both support and interdependence links, as described in the legend. It
also shows a comparison with the empirical percolation profile of
networks realized from this ensemble, which is in excellent agreement
with the theoretical prediction.

\begin{figure}[ht!]
  \setlength{\unitlength}{0.95\columnwidth}
  \drawbox{
    \begin{picture}(1,0.8)
      \put(-0.03,0.4){\includegraphics[width=0.28\textwidth, trim=0 0.5cm 0 0]{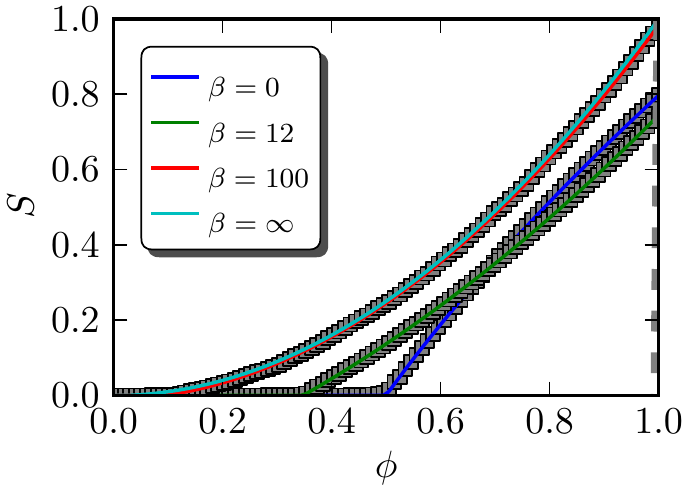}}
      \put(-0.02,0.4){(a)}

      \put(0.58, 0.42){\begin{minipage}[b]{0.095\textwidth}
          \centering
          \includegraphics[trim=0.3cm 0.55cm 0.5cm 0.4cm, clip=true, width=\textwidth]{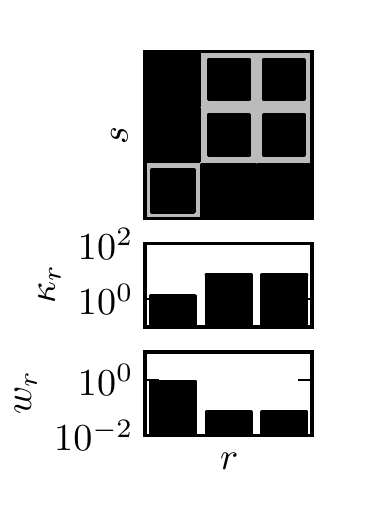}\\
          \smaller $\qquad\beta = 12$
      \end{minipage}}
      \put(0.80, 0.42){\begin{minipage}[b]{0.095\textwidth}
          \centering
          \includegraphics[trim=0.3cm 0.55cm 0.5cm 0.4cm, clip=true, width=\textwidth]{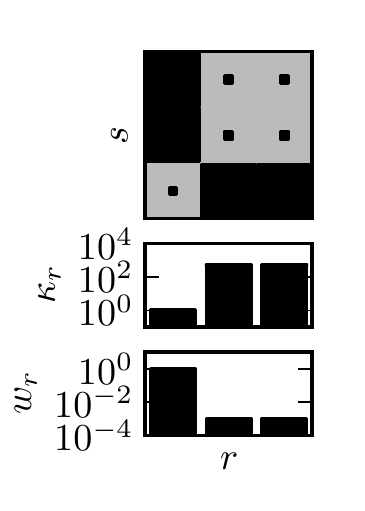}\\
          \smaller $\qquad\beta = 100$
      \end{minipage}}
      \put(0.59, 0.4){(b)}

      \put(-0.01, 0.0){
        \drawbox{\begin{minipage}[b]{0.13\textwidth}
          \centering
          \includegraphics[width=\textwidth,trim=0.8cm 0.0cm 0.5cm 0.0cm]{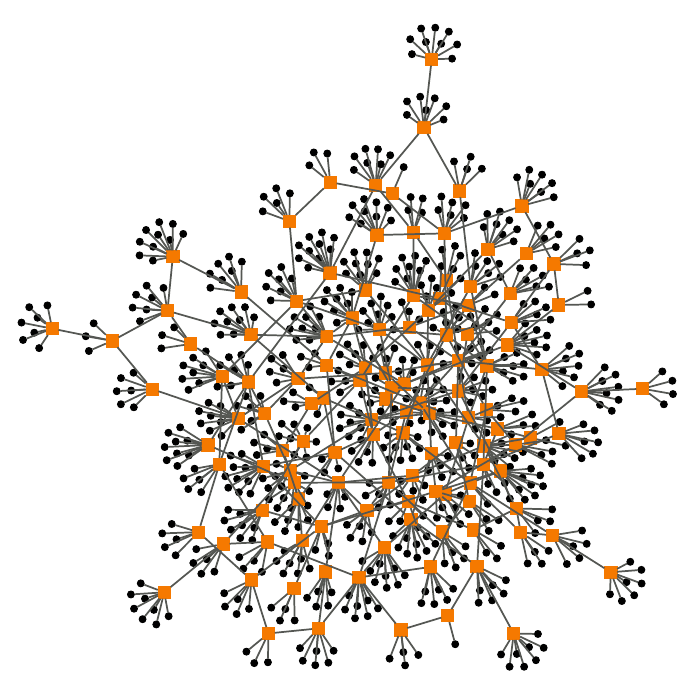}\\
         \smaller $\beta=10$
        \end{minipage}}
      }
      \put(0.34, 0.0){
        \drawbox{\begin{minipage}[b]{0.13\textwidth}
          \centering
          \includegraphics[width=\textwidth,trim=0.8cm 0.0cm 0.5cm 0.0cm]{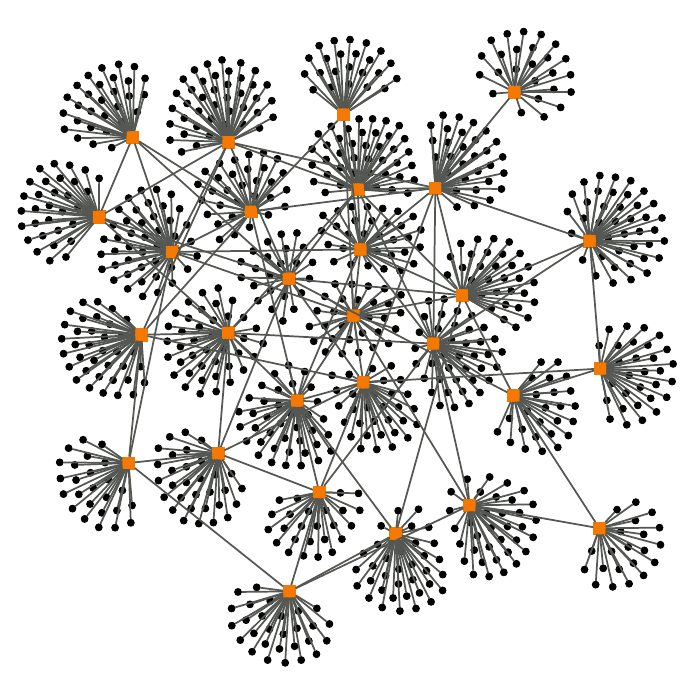}\\
         \smaller $\beta=20$
        \end{minipage}}
      }
      \put(0.7, 0.0){
        \drawbox{\begin{minipage}[b]{0.13\textwidth}
          \centering
          \includegraphics[width=\textwidth,trim=0.6cm 0.0cm 0.5cm 0.0cm]{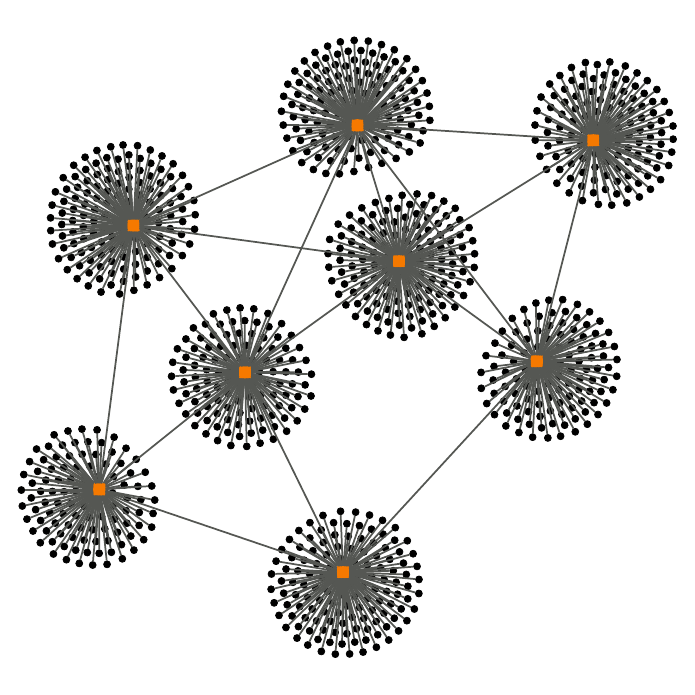}\\
         \smaller $\beta=40$
        \end{minipage}}
      }
      \put(0.0, 0.0){(c)}
    \end{picture}}
  \caption{\label{fig:opt-aut}(Color online) (a) $S$ as a function of
  $\phi$, for random failures, different $\beta$ values, $\avg{k}=2$ and
  $\hat{\avg{k}}=0$. The solid lines are results obtained via
  Eqs.~\ref{eq:mf_u} and~\ref{eq:mf_phi}, and the filled symbols are
  obtained empirically with a network realization of size $N=10^6$. The
  dashed curve close to $\phi=1$ corresponds to targeted attacks for
  $\beta\to\infty$; (b) Correlation matrix $e_{rs}/e_re_s$ (top), block
  sizes $w_r$ and average degrees $\kappa_r$, for different $\beta$
  values; (c) Ensemble samples with $\avg{k}=2$ and $N=10^3$ nodes for
  different $\beta$ values. The orange (square) nodes belong to the core
  block.}

\end{figure}
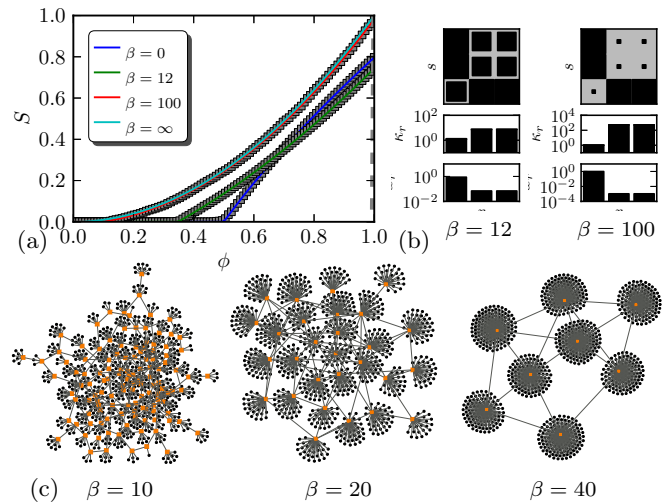

We are interested in obtaining the block topologies for which the
robustness against failure or attack are
optimal. Following~\cite{schneider_mitigation_2011} we will consider the
total robustness of a network ensemble to be given by
\begin{equation}\label{eq:R}
  R = 2\int_0^1S(\phi)d\phi,
\end{equation}
where the factor $2$ is chosen so that $R \in [0,1]$, with $R=1$ being
possible only in the hypothetical scenario $S(\phi)=\phi$, attainable
only for infinitely dense networks. Thus our ultimate task is to find
the parametrization of the blockmodel ensemble which maximizes
Eq.~\ref{eq:R}, under suitable constraints. In the following, we focus
on the special case with uniform failure of nodes within a single block,
$\phi^r_k \equiv \phi_r$; hence $f_0^r(z)=\phi_rg_0^r(z)$.  The total
dilution becomes simply $\phi = \sum_rw_r\phi_r$.  [Without loss of
generality, since heterogeneous degree-based dilution can still be
achieved if nodes of different degrees belong implicitly to different
blocks.] In the case of random failure we have simply, $\phi_r =
\phi$. In the case of targeted attacks we will use $\phi_r \propto
e^{\kappa_r (1 - b)/b}$, where $b \in [0, 1]$ must be so chosen to
achieve a desired total $\phi$.

Instead of directly finding the network topology parametrization which
maximizes $R$, we proceed by obtaining the configuration which delivers
a specified value of $R$, and is otherwise maximally random. We do so in
order to identify network structures with a varied degree of robustness,
and to isolate the most fundamental characteristic responsible for its
increase. Hence, we consider a \emph{null model} of robust networks,
where superfluous characteristics are explicitly discarded. More
specifically, we maximize the entropy of network
ensemble~\cite{bianconi_entropy_2009}, subject to the constraint that
the average robustness is fixed~\cite{park_statistical_2004}. This
amounts to finding the critical points of the Lagrangian $\Lambda =
\Sigma - \tilde{\beta}(R-R^*)$, where $\Sigma=\ln\Omega$ is the
microcanonical entropy (with $\Omega$ being the number of different
network realizations for a specific ensemble parametrization), $R$ and
$R^*$ are the actual and desired average robustness respectively, and
$\tilde{\beta}$ is a Lagrange multiplier. Instead of fixing $R^*$, one
may fix $\tilde{\beta}$, and this becomes equivalent to minimizing the
\emph{free energy} of the ensemble,
\begin{equation}\label{eq:fe}
  \mathcal{F} = -NR - \Sigma/\beta,
\end{equation}
since $\Lambda = \tilde{\beta}(\mathcal{F}/N + R^*)$, and $\beta \equiv
-\tilde{\beta}/N$.  In Eq.~\ref{eq:fe}, the value $-NR$ plays the role
of average energy, and $\beta$ is the inverse temperature, which can
also be interpreted as a ``selective
pressure''~\cite{peixoto_emergence_2012}, since for $\beta=0$ the
entropy is strictly maximized (i.e. all networks occur with equal
probability), and conversely for $\beta\to\infty$ the average robustness
is strictly maximized. In our case we have $\Sigma=\mathcal{S} +
\hat{\mathcal{S}}$, where $\mathcal{S}$ is the stochastic blockmodel
entropy~\cite{peixoto_entropy_2012},
\begin{equation}\label{eq:S}
  \mathcal{S} =-\sum_rn_r\sum_kp^r_k(\ln p^r_k + \ln k!) -
  \frac{1}{2}\sum_{rs}e_{rs}\ln\left(\frac{e_{rs}}{e_re_s}\right),
\end{equation}
which includes the entropy of the degree distributions of the individual
blocks, and $\hat{\mathcal{S}}$ is equivalently defined as a function of
$\hat{e}_{rs}$ and $p^r_{\hat{k}}$~\footnote{Eq.~\ref{eq:S} omits
constant terms which are not relevant to the optimization parameters,
and is valid in the sufficiently sparse case
(see~\cite{peixoto_entropy_2012}), which is also assumed for
Eqs.~\ref{eq:mf_u} and~\ref{eq:mf_phi}.}.

Solving the system given by Eqs.~\ref{eq:mf_u} and~\ref{eq:mf_phi}
cannot be performed analytically, and thus the same holds for obtaining
$R$ and $\mathcal{F}$. Hence we must resort to solving
Eqs.~\ref{eq:mf_u} and~\ref{eq:mf_phi} numerically (by simple
iteration), in order to obtain Eqs.~\ref{eq:R} and~\ref{eq:fe}. The
minimization of Eq.~\ref{eq:fe} can then be performed within arbitrary
precision with any suitable minimization algorithm (see
\emph{Supplemental Material} for details).

The ensemble can have additional constraints, which must be imposed when
minimizing $\mathcal{F}$. One such constraint we will consider
throughout this paper is that the average degree $\avg{k}$ will be fixed
at a given value, which represents the putative cost of adding extra
edges to the network, compared to simply replacing them. Furthermore, in
order to restrict the number of degrees of freedom of the model to a
manageable size, we will assume that all blocks have the same type of
degree distribution, which can be different only in their average value,
$\kappa_r$. This does not alter very strongly the type of networks which
are ultimately attainable, since in principle one can construct many
different topologies by composing a larger number of such blocks. We
will also forbid nodes with degree zero, since these never belong to a
macroscopic component. The degree distribution we use is a modified
Poisson distribution $p^r_k = (1-\delta_{k,0})\kappa^k_r / (e^{\kappa_r}
- 1) k!$ [identical to a regular Poisson, but with support $k\ge1$],
which is generated by $g^r_0(z) = z e^{(\kappa_r - 1) (z - 1)}$, which
of course implies that $\kappa_r\ge 1$ for all $r$. The same
distribution will also be used for the interdependence degrees,
$p^r_{\hat{k}}$. Note that this simplifies the entropy to $\mathcal{S} =
\sum_rn_r\ln(e^{\kappa_r}-1) -
\frac{1}{2}\sum_{rs}e_{rs}\ln(e_{rs}/n_rn_s)$, and equivalently for
$\hat{\mathcal{S}}$. Finally, the total number of blocks $B$ will also
remain fixed during the minimization of $\mathcal{F}$.

\begin{figure}[hb!]
  \setlength{\unitlength}{0.95\columnwidth}
  \drawbox{
    \begin{picture}(1,0.3)
      \put(-0.028,0.0){\includegraphics[width=0.43\columnwidth, trim=0 0.5cm 0 0]{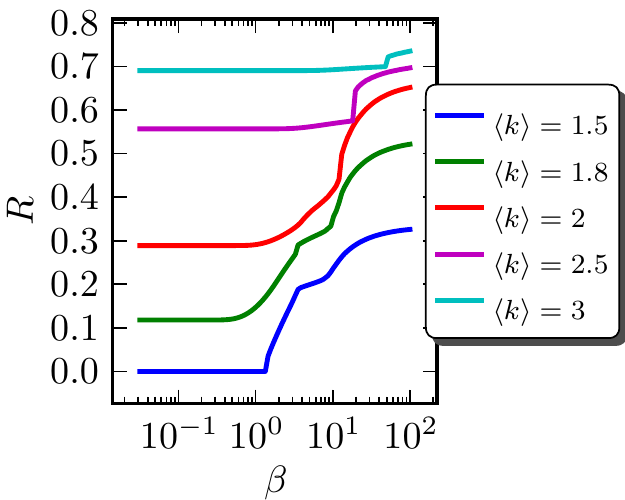}}
      \put(-0.03,0.0){(a)}

      \put(0.425, 0.0){\includegraphics[width=0.134\textwidth, trim=0 0.7cm 0 0]{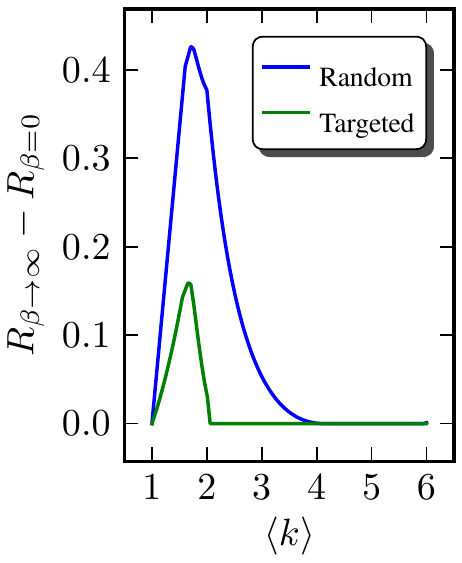}}
      \put(0.425, 0.0){(b)}

      \put(0.715,0.){\includegraphics[width=0.15\textwidth, trim=0 0.7cm 0 0]{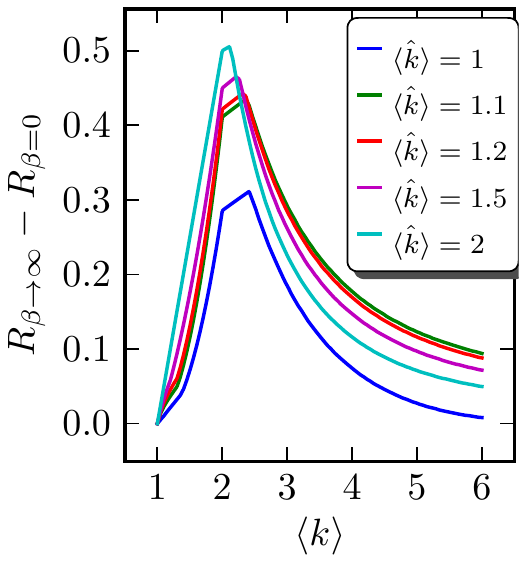}}
      \put(0.72,0.){(c)}

    \end{picture}
  }
  \caption{\label{fig:opt-limit}(Color online) (a) Robustness $R$
  (Eq.~\ref{eq:R}) as a function of $\beta$ for different average
  degrees $\avg{k}$, with $\hat{\avg{k}}=0$; (b, c) Robustness
  difference between optimal ensembles ($\beta\to\infty$) and random
  networks ($\beta=0$) with the same $\avg{k}$, for (b) $\hat{\avg{k}} =
  0$ and (c) $\hat{\avg{k}} \ge 1$.}
\end{figure}
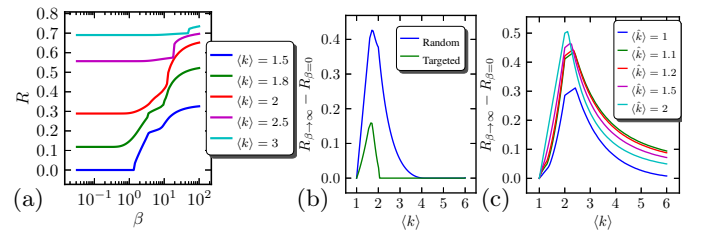

\begin{figure}[htb!]
  \setlength{\unitlength}{\columnwidth}
  \drawbox{
    \begin{picture}(1, 0.35)
      \put(-0.03,0.0){\includegraphics[width=0.2\textwidth, trim=0 0.5cm 0 0]{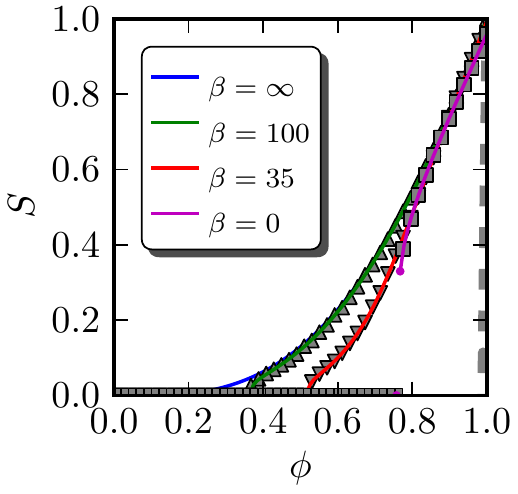}}
      \put(-0.02,0.0){(a)}

      \put(0.4, 0.){\begin{minipage}[b]{0.095\textwidth}
          \centering
          \includegraphics[trim=0.15cm 0.2cm 0.5cm 0.4cm, clip=true, width=\textwidth]{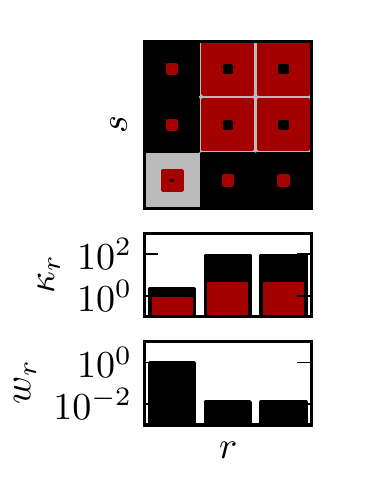}\\
          \smaller $\qquad\beta = 10^2$
      \end{minipage}}
      \put(0.4, 0.){(b)}

      \put(0.66, 0.0){
        \drawbox{\begin{minipage}[b]{0.13\textwidth}
          \centering
          \includegraphics[width=\textwidth,trim=0.5cm 0.0cm 0.5cm 0.0cm]{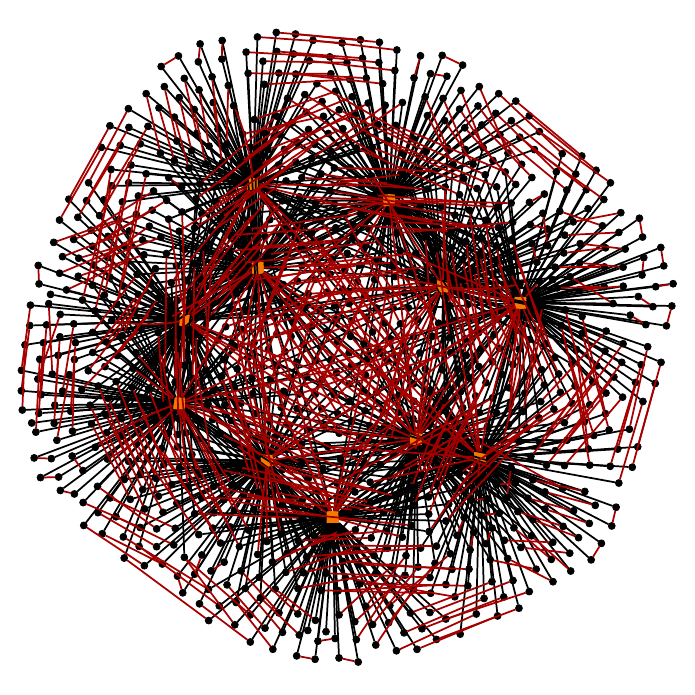}\\
         \smaller $\beta=30$
        \end{minipage}}
      }
      \put(0.66, 0.0){(c)}
    \end{picture}
  }

  \caption{\label{fig:opt-inter}(Color online) (a) Same as
  Fig.~\ref{fig:opt-aut}a, with $\avg{k}=4$, $\hat{\avg{k}}=1.1$.;
  (b) Same as Fig.~\ref{fig:opt-aut}b, and the same constraints as in
  (a). The correlation matrix $\hat{e}_{rs}/\hat{e}_r\hat{e}_s$ and
  average interdependence degrees $\hat{\kappa}_r$ are shown in red; (c)
  Same as Fig.~\ref{fig:opt-aut}a, with $\avg{k}=2$, $\hat{\avg{k}}=1.1$
  and $N=10^4$. The red edges are interdependence links.}
\end{figure}
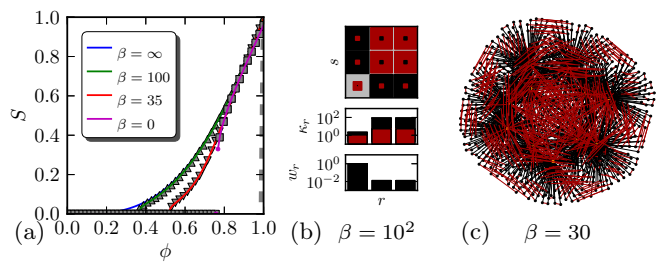

In absence of any further constraints, we start with the case of random
failures, and no interdependence links ($\hat{\kappa}_r=0$). We observe
that increasing $\beta$ leads from a fully random graph ($\beta=0$) to a
topology which can be significantly more robust (see
Fig.~\ref{fig:opt-aut}a). For $B=2$ blocks, this topology is a
core-periphery (CP) structure~\cite{holme_core-periphery_2005,
borgatti_models_2000}, where one block is much smaller and has a much
larger average degree, $\kappa_r$, and attracts a large portion of the
edges. One might expect this to be a special case when one imposes
$B=2$, and that different robust topologies may arise for
$B>2$. However, this turns out not to be the case, and we find
\emph{exactly} the same CP topology for any value of $B$ larger than
two, in the sense that two or more blocks may be merged together, with
the ensemble remaining equivalent, until only $B=2$ blocks are left. In
Fig.~\ref{fig:opt-aut}b are shown two resulting ensembles for different
$\beta$ and with $B=3$, for $\avg{k}=2$. The size and average degree of
the core block become smaller and larger, respectively, for larger
values of $\beta$. Examples of how such networks can look like are shown
in Fig.~\ref{fig:opt-aut}c. For sufficiently large $\beta$, the vast
majority of periphery nodes are connected exclusively to core nodes,
which are much smaller in number (infinitesimally so for
$\beta\to\infty$), and which are connected among themselves, forming a
\emph{central backbone}~\footnote{This is reminiscent of the random
graphs with bimodal degree distribution in~\cite{valente_two-peak_2004,
paul_optimization_2004, tanizawa_optimization_2005}.  However, these do
not possess a CP topology, since all nodes are forced to be randomly
connected.}.

Observing how the value of $R$ changes with $\beta$ exposes a structural
phase transition at a critical value of $\beta=\beta^*$, below which the
networks are fully random ($e_{rs} \propto e_{r}e_{s}$ and
$\kappa_r=\avg{k}$), and above which the CP topology is observed (see
Fig~\ref{fig:opt-limit}a). The value of $\beta^*$ becomes smaller for
$\avg{k}$ close to two, and larger for either sparser or denser
graphs. The CP topology is particularly advantageous for sparse graphs,
with lower $\avg{k}$, but is less so for denser graphs, as can be seen
in Fig.~\ref{fig:opt-limit}a, which shows a comparison between the
$\beta=0$ (random) and $\beta=\infty$ (optimal) cases, as a function of
$\avg{k}$.

When node interdependence is enforced ($\hat{\kappa}_r \ge 1$
~\footnote{We could also allow autonomous nodes,
  i.e. $\hat{\kappa}_r < 1$~\cite{schneider_towards_2011}. But allowing
  this leads to trivial block structures where all interdependence edges
  are confined to an isolated block of insignificant size. We therefore
  focus on the harder case where autonomous nodes are not
  allowed.}), the same CP structure emerges for larger values of
$\beta$, as shown in Fig.~\ref{fig:opt-inter}b. For $\beta\to\infty$,
the percolation transition changes from abrupt to continuous (see
Fig.~\ref{fig:opt-inter}a). The average interdependence degrees,
$\hat{\kappa}_r$, become correlated with the connectivity degrees,
$\kappa_r$, with most interdependence edges lying between two core
nodes, which therefore benefit more from the extra redundancy, to the
indirect benefit of the periphery nodes as well. In comparison to the
$\hat{\avg{k}}=0$ case, denser networks benefit more from the CP
topology, since it removes the abrupt phase transition. The benefit
diminishes for larger $\hat{\avg{k}}$, since the larger redundancy
already provides an inherent robustness to random networks. A special
case is $\hat{\avg{k}}=1$, which shows a lower improvement, because it
strictly forbids intra-core redundancy.

As seen in Figs.~\ref{fig:opt-aut}a and~\ref{fig:opt-inter}a, the CP
topology is excellent against random failure, but is extremely fragile
against targeted attacks, in which the core nodes are removed before the
rest (the value of $S$ vanishes for an infinitesimal value of
$1-\phi$). Instead, if one optimizes $R$ against targeted attacks, one
obtains that the random block topology is the most robust. The only
exceptions are very sparse networks with $\avg{k} \lesssim 2$ (see
Fig.~\ref{fig:opt-limit}b). Since these networks are close or below the
percolation threshold even for $\phi=1$, the CP topology is still an
improvement over the fully random case, even if gets destroyed very
quickly by a targeted attack.

\begin{figure}
  \setlength{\unitlength}{0.95\columnwidth}
  \drawbox{
    \begin{picture}(1,0.75)
      \put(0.03, 0.0){\includegraphics[width=0.21\textwidth, trim=0 0.5cm 0 0]{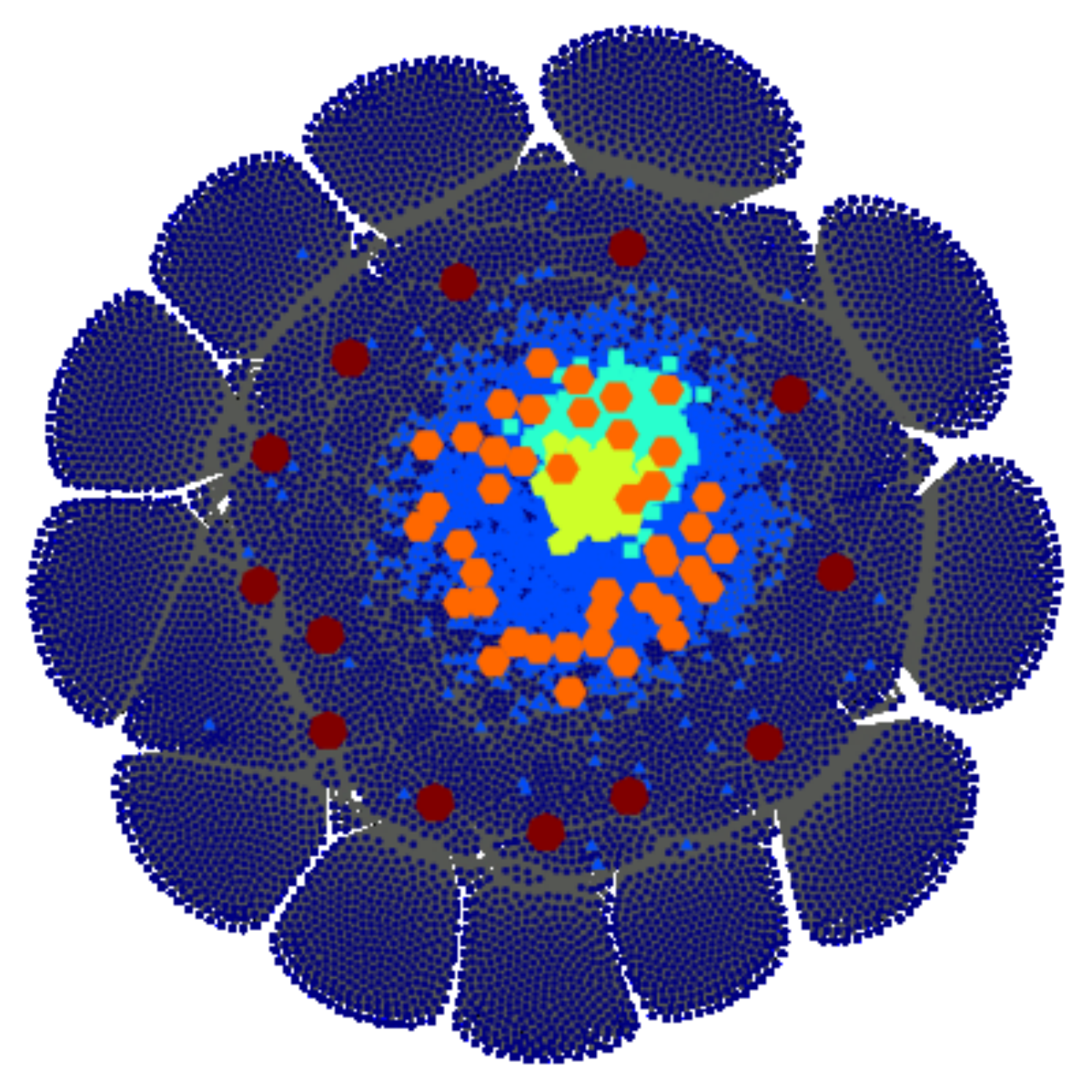}}

      \put(0.53, 0.0){\includegraphics[width=0.21\textwidth, trim=1cm 2cm 0 0]{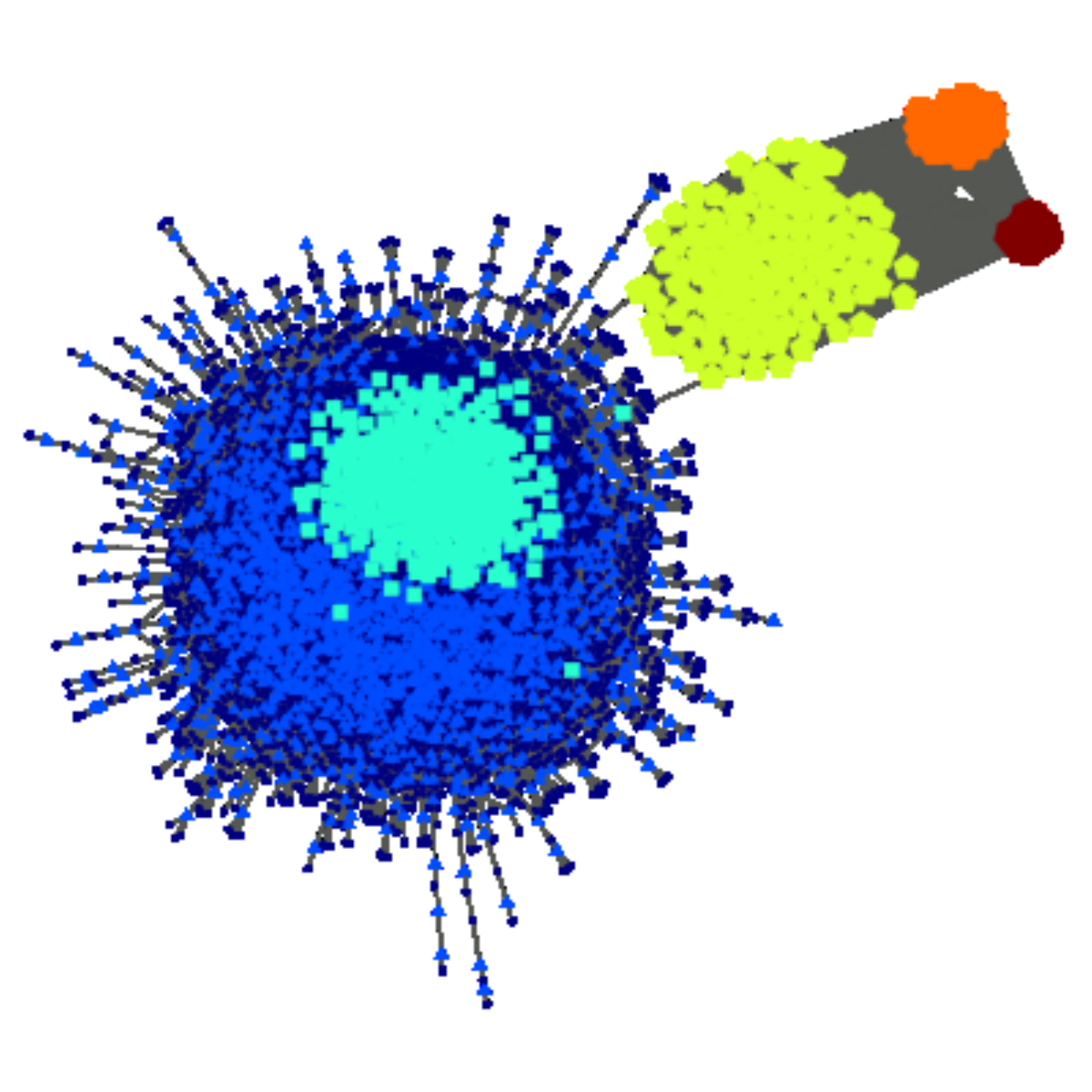}}

      \put(-0.03,0.45){\includegraphics[width=0.125\textwidth, trim=0 0.5cm 0 0]{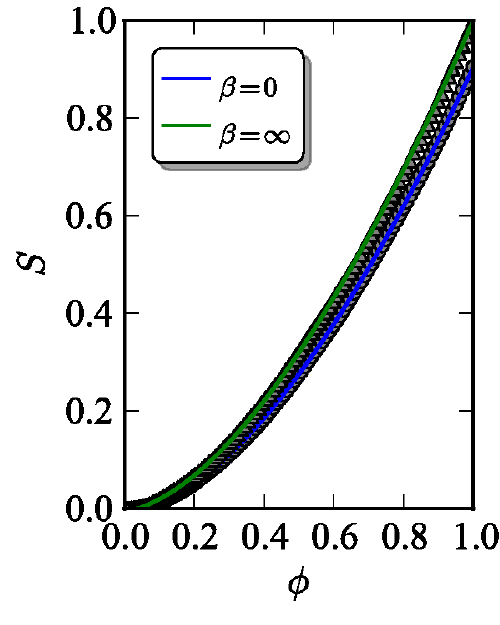}}
      \put(0.255, 0.45){\includegraphics[width=0.12\textwidth, trim=0.35cm 0.5cm 0 0, clip=true]{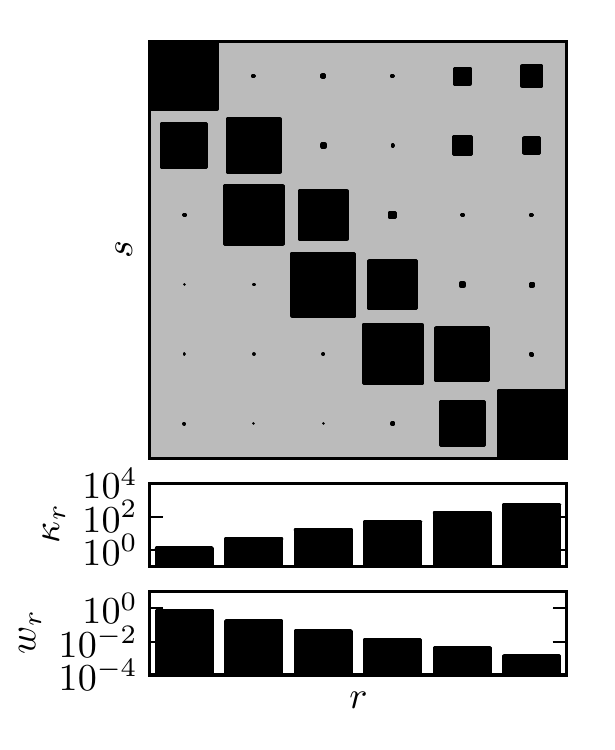}}

      \put(0.5, 0.45){\includegraphics[width=0.125\textwidth, trim=0 0.5cm 0 0]{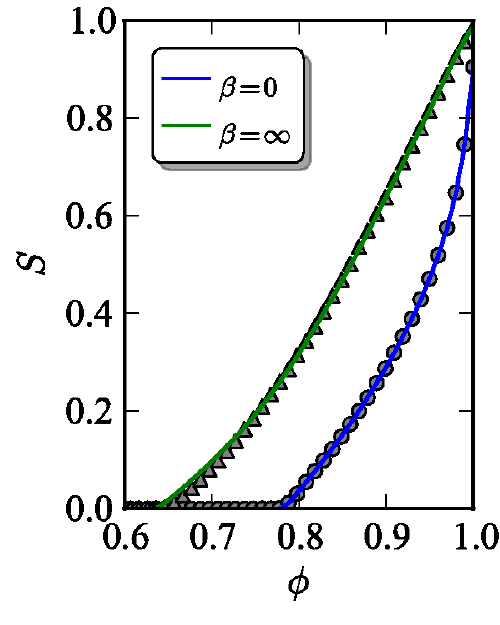}}
      \put(0.77, 0.45){\includegraphics[width=0.12\textwidth, trim=0.35cm 0.5cm 0 0, clip=true]{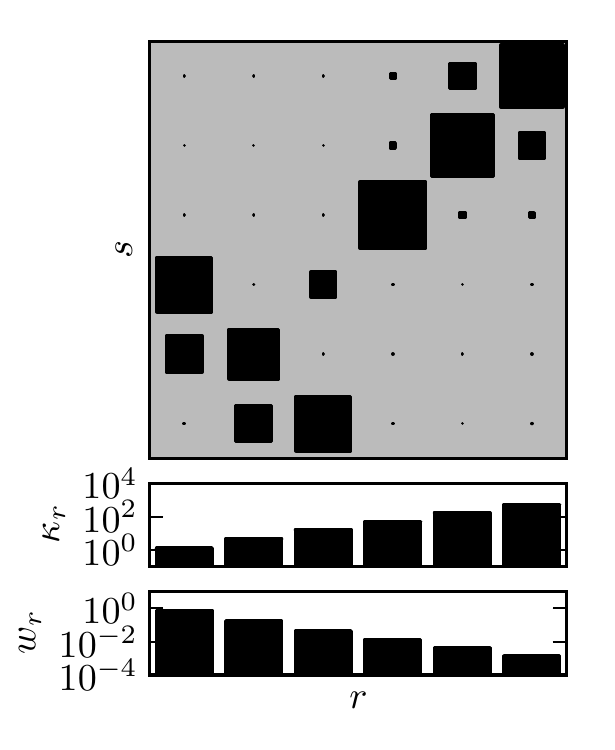}}

      \put(-0.01,0.42){(a)}
      \put(0.5,0.42){(b)}
      \put(0.0, 0.0){(c)}
      \put(0.51, 0.0){(d)}
    \end{picture}}

  \caption{\label{fig:opt-fixed}(Color online) (a, b) Same as
  Fig.~\ref{fig:opt-aut}a for (a) random failures and (b) targeted attacks. In both cases the
  values $w_r$ and $\kappa_r$ are kept fixed. (c, d) Ensemble samples
  with $N=10^4$ nodes, $\beta\to\infty$, for (c) random failures and
  (d) targeted attacks.}
\end{figure}
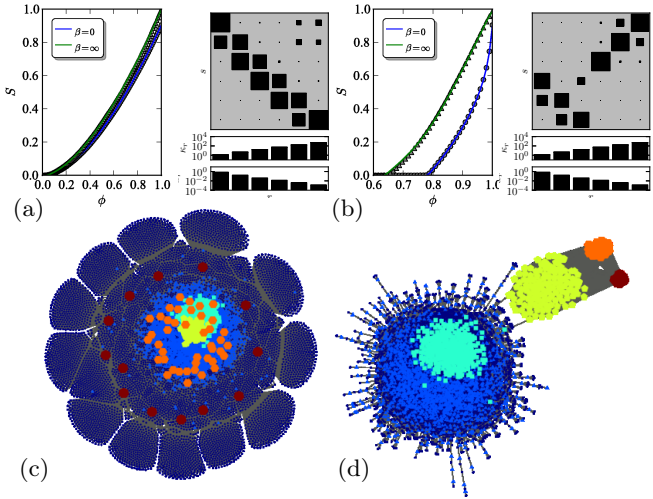

In some realistic situations there are degree constraints restricting
the accessible topologies~\cite{schneider_mitigation_2011}. Here we
introduce this by forcing that the blocks have prescribed sizes and
average degrees (see \emph{Supplemental material} for details).  As can
be seen in Fig.~\ref{fig:opt-fixed}a, in the case of random failure, the
optimal topology approaches the CP configuration as much as the imposed
constraints allow, and the inter-block connectivity becomes strongly
dissortative, with the nodes with largest degrees serving as
intermediaries between the nodes with the lowest degrees and an
innermost ``core'' composed of more of such alternating layers. For
targeted attacks the resulting topology is quite different: The blocks
with higher $\kappa_r$ are ``expelled'' from the network, since they are
the first to be removed in an attack. The resulting topology is highly
assortative, with blocks with similar $\kappa_r$ connecting
preferentially to themselves. This is qualitatively equivalent to the
``onionlike'' topology found empirically
in~\cite{schneider_mitigation_2011}. An exception to this are the blocks
with small $\kappa_r$. These do form a CP structure, since an
assortative connectivity would have very weak percolation properties
even when no nodes are removed, i.e. $\phi=1$ (similarly to the case
$\avg{k} \lesssim 2$ discussed previously). Very similar results are
obtained when interdependence is introduced (see \emph{Supplemental
material}).

The topologies described above arise when no constraint other than those
considered is being imposed, and the ensemble entropy is maximized.
This is not often the case for real systems, since they may be subject
to competing selective pressures, with robustness being only one of
them, as well as other topological restrictions, frozen accidents,
etc. Nevertheless, the analysis presented here suggests that a simple
core-periphery structure is the most natural configuration which
achieves robustness against random failure (and that a random,
non-centralized topology works best against targeted
attacks). Therefore, it not a surprise that a CP topology is indeed
found to some extent in many real
systems~\cite{holme_core-periphery_2005}, such as the
Internet~\cite{doyle_robust_2005}, and gene regulation
networks~\cite{peixoto_emergence_2012}. Also, our results indicate that
other features found in real systems, such as scale-free degree
distributions, are not a strict necessity for robustness, and do not
arise naturally for this purpose, and thus may be simply a byproduct of
a non-equilibrium organization process.

\bibliographystyle{apsrev4-1}
\bibliography{bib}

\pagebreak

\begin{widetext}


\section*{Supplementary material}

# Supplemental Material: Evolution of robust network topologies

Tiago P. Peixoto[*] and Stefan Bornholdt[†]

*Institut für Theoretische Physik, Universität Bremen, Hochschulring 18, D-28359 Bremen, Germany*

## GENERATING FUNCTION PERCOLATION FORMALISM

Following [1–3], we derive self-consistency equations for the percolation properties of stochastic blockmodel ensembles as follows. If we let $u_r$ describe the probability that a node belonging to block $r$ is not in a macroscopic component via one of its neighbors, the dilution variable $\phi_k^r$ as the fraction of nodes belonging to block $r$ and with degree $k$ which are not removed from the network, and additionally the *interdependence dilution* $\hat{\phi}_r$ defined as the fraction of nodes from block $r$ which were not removed due to the failure of the support nodes, we can write,

$$u_r = \sum_s m_{rs} \sum_k (1 - \hat{\phi}_s \phi_{k+1}^s + \hat{\phi}_s \phi_{k+1}^s u_s^k) q_k^s, \qquad (1)$$

where $m_{rs} = e_{rs}/n_r \kappa_r$ is the asymmetric matrix defining the fraction of edges adjacent to vertices of block $r$ which are also adjacent to block $s$, and $q_k^r = p_{k+1}^r (k+1)/\kappa_r$ is the excess degree distribution of block $r$, and $\kappa_r = \sum_s e_{rs}/n_r$ is the average degree of block $r$. This can be rewritten in compact form as

$$u_r = \sum_s m_{rs} \left[ 1 - \hat{\phi}_s f_1^s(1) + \hat{\phi}_s f_1^s(u_s) \right], \qquad (2)$$

where $f_1^r(z) = \sum_k q_k^r \phi_{k+1}^r z^k$ is the generating function of the excess degree distribution. Note that we have $f_1^r(z) = f_0^{r\prime}(z)/f_0^{r\prime}(1)$ where $g_0^r = \sum_k p_k^r z^k$ and $f_0^r(z) = \sum_k p_k^r \phi_k^r z^k$ are the generating functions of the original and diluted degree distributions, respectively. The fraction of nodes of block $r$ which belong to a macroscopic component is given by $S_r = \hat{\phi}_r S_r^0$, where $S_r^0 = f_0^r(1) - f_0^r(u_r)$. The total fraction of nodes which belong to a macroscopic component is simply $S = \sum_r w_r S_r$, where $w_r = n_r/N$ is the fraction of nodes belonging to block $r$, and the total dilution in the network is likewise $\phi = \sum_{r,k} w_r p_k^r \phi_k^r$.

In the case of no interdependence edges, we have simply $\hat{\phi}_s = 1$, and Eq. 2 describes $S_r$ completely. Otherwise, an additional fraction of nodes will be removed from the macroscopic components, if the nodes to which they depend do not belong to the component. The fraction $\hat{\phi}_r$ of nodes which are not removed in this way can be obtained by

$$\hat{\phi}_r = \sum_{\hat{k}} p_{\hat{k}}^r \left[ 1 - (1 - \sum_s \hat{m}_{rs} S_s^0)^{\hat{k}} \right] + p_0^r, \qquad (3)$$

where $\hat{k}$ are the interdependence degrees, and $p_{\hat{k}}^r$ is the interdependence degree distribution of block $r$. Here it is assumed that the interdependence is undirected (if node $u$ is dependent on $v$, then $v$ is also dependent on $u$), and that at least one dependency must not fail in order for the dependent node not to fail (i.e. redundant interdependence), and furthermore that a node with $\hat{k} = 0$ is autonomous (see below for variations of these rules). We may write this equation in a more compact form as,

$$\hat{\phi}_r = 1 - \hat{f}_0^r(1 - \sum_s \hat{m}_{rs} S_s^0) + \hat{f}_0^r(0), \qquad (4)$$

where $\hat{f}_0^r(z) = \sum_{\hat{k}} p_{\hat{k}}^r z^{\hat{k}}$ is the interdependence degree generating function. Note that the right hand side of Eq. 4 depends only on $u_r$ variables (via $S_r^0$), and thus may be included directly into Eq. 2, resulting in a total of only $B$ equations. Similarly to [3], here we omit the detailed description of the failure cascades as done in e.g. [2], however these correspond simply to repeated iterations of Eqs. 2 and 4, where one must first obtain the convergence of the $u_r$ variables, and then of the $\hat{\phi}_r$ variables, in succession. However, if one is only interested in the steady state, simultaneously iterating Eqs. 2 and 4 is more straightforward.

One may also consider different interdependence rules by slightly modifying Eq. 4. For instance, in the case of non-redundant interdependence, where *all* dependencies are essential, so that if any one of them fails, the dependent node fails as well, can be described by replacing Eq. 4 by

$$\hat{\phi}_r = \hat{f}_0^r(\sum_s \hat{m}_{rs} S_s^0) + \hat{f}_0^r(0). \qquad (5)$$

Furthermore, one may also consider the situation where the interdependence is directed, i.e. if node $u$ depends on node $v$, $v$ does not necessarily depend on $u$ (in this situation it assumed that $\hat{e}_{rs}$ is also directed, and thus asymmetric, and the variables $\hat{k}$ describe the dependence *in-degree*). This can be implemented by an almost trivial modification, simply by replacing $S_s^0 \to S_s$ in Eq. 4 or 5, where we have, as before, $S_r = \hat{\phi}_r S_r^0$. This is simply because a node will fail either if it does not belong to a macroscopic component, or if its dependence neighbors fail. However, in the case of directed interdependence, the likelihood that two nodes are mutually interdependent is vanishingly small, and thus the support node will depend exclusively on other nodes, which will themselves not fail with probability given by $\hat{\phi}_r$. This introduces a direct coupling between the variables $\hat{\phi}_r$ (which is only indirect in the undirected case). Although seemingly innocuous, this modification can drastically change the percolation properties of the system, in particular when the

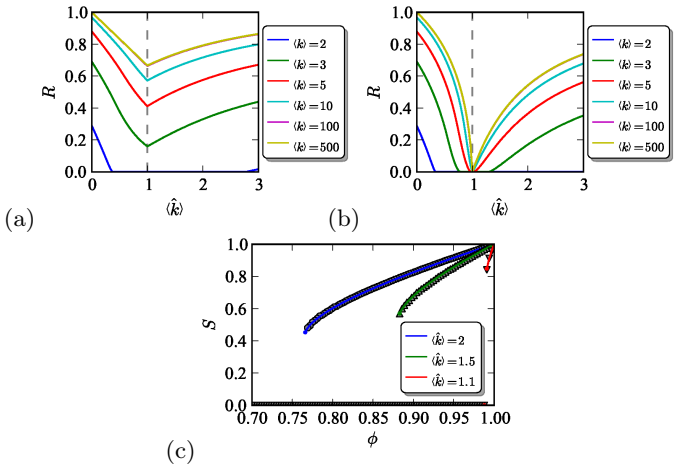

FIG. 1. Robustness $R$ to uniform node removal of random networks with one block $B = 1$, as a function of $\langle \hat{k} \rangle$, and different values of $\langle k \rangle$, for (a) undirected interdependence and (b) directed dependence; (c) Relative size of the largest component $S$ as a function of the dilution $\phi$, for a random graph with $\langle k \rangle = 5$, directed dependencies and different $\langle \hat{k} \rangle$ values. The solid lines are results obtained via Eqs. 2 and 4 (with $S_s^0 \to S_s$), and the filled symbols are obtained empirically with a network realization with $N = 10^6$ nodes.

average degrees $\hat{\kappa}_r$ are close to one. Indeed if all nodes have $\hat{k} = 1$, the failure of an *infinitesimal fraction* of nodes will *always* cause a global cascade, independently of how well connected the network is. This is easy to understand, since the removal of a single node will necessarily cause the failure of another node, which will trigger the failure of a third one, and so on, until a loop is reached. The average size of these cascades diverges, and thus the values of $S_r$ will drop to zero. The robustness of random networks (B=1) with different values of $\langle \hat{k} \rangle$ and $\langle k \rangle$ are shown in Fig. 1, for both connectivity and interdependence degree distributions given in the main text if $\langle k \rangle \geq 1$ or $p_k = \langle k \rangle \delta_{k,1} + (1 - \langle k \rangle)\delta_{k,0}$ if $\langle k \rangle < 1$. For undirected interdependence, we have that $R$ approaches one for $\langle k \rangle \to \infty$, for all values of $\langle \hat{k} \rangle$. However for directed dependence, we have $R$ dropping to zero as $\langle \hat{k} \rangle$ approaches one, independently of $\langle k \rangle$.

In all cases, the percolation threshold can be obtained by examining the Jacobian $J = \{ \partial (u_r' | \hat{\phi}_r') / \partial (u_s | \hat{\phi}_s) \}$, where $u_r'$ and $\hat{\phi}_r'$ correspond to the right-hand side of Eqs. 2 and 4, respectively. If the largest eigenvalue $\lambda$ of $J$ at the fixed point $u_r = 1$ and $\hat{\phi}_r = \hat{f}_0^r(0)$ exceeds one, this solution is unstable, and thus some value of $S_r$ must be finite, and hence the condition $\lambda = 1$ marks the percolation transition.

## MINIMIZATION OF THE FREE ENERGY, $\mathcal{F}$

As mentioned in the main text, the computation of $R$ and thus $\mathcal{F}$ must be done numerically, by iterating Eqs. 2 and 4 until convergence within some desired precision is reached. This means that the minimization of $\mathcal{F}$ must also be performed numerically. We used different algorithms, depending on the set of constraints used. In the case of $\langle \hat{k} \rangle = 0$, the L-BFGS-B quasi-Newton method [4], with gradient obtained with finite differences, delivered fast and reliable results if the number of degrees of freedoms were not very large ($B = 2$ or $3$). For larger number of blocks, the non-gradient COBYLA algorithm [5] performed better. In the case $\langle \hat{k} \rangle > 0$, the discontinuity in $S(\phi)$ makes the gradient estimation unreliable, and the COBYLA algorithm is a better choice. In both cases, when improved precision was necessary, a repeated sequential one-dimensional minimization over all variables using Brent's method [6] was performed, at the expense of longer computational time. In the case of very large values of $B$, an initial estimation using the ALPSO stochastic swarm-based algorithm [7] was used, which was then improved upon, whenever possible, with COBYLA.

In all cases, the constrained optimization problem was transformed into an unconstrained one, via appropriate transformations. The normalization constraints, $\sum_r w_r = 1$ and $\sum_{rs} e_{rs} = N\langle k \rangle$ can be imposed simply by transforming $w_r = C \exp(\tilde{w}_r)$, and $e_{rs} = D \exp(\tilde{e}_{rs})$ with $C^{-1} = \sum_r w_r$ and $D^{-1} = \sum_{rs} e_{rs}/N\langle k \rangle$, and solving for $\tilde{w}_r$ and $\tilde{e}_{rs}$. Note that we can set, e.g. $\tilde{w}_0 = 0$ and $\tilde{e}_{00} = 0$ so that only $2B + B(B-1)/2 - 2$ free variables remain. The degree constraints $\kappa_r \geq 1$ can be imposed via the transformations $\kappa_r = \tilde{\kappa}_r$ if $\kappa_{\min} \equiv \min(\{\kappa_r\}) \geq 1$ otherwise $\kappa_r = a_r \tilde{\kappa}_r + 1$ where $a_r = (\kappa_r - 1)/\sum_s w_s(\kappa_s - \kappa_{\min})$. The new row and column sums of $e_{rs}$ given by $\sum_s e_{rs} = \sum_s e_{sr} = N w_r \kappa_r$ can then be enforced in the $e_{rs}$ matrix via a Sinkhorn scaling [8]. The same transformations can be used for the $\hat{e}_{rs}$ matrix. With all these transforms in place, the optimization problem becomes fully unconstrained, with $\tilde{w}_r$, $\tilde{e}_{rs}$ and $\hat{e}_{rs}$ lying in the interval $[-\infty, \infty]$. Note that the variable $N$ is only included in the text to slightly simplify the notation, but it is entirely unnecessary during the optimization, since one needs to work only with normalized variables.

In all minimizations performed, the same solutions were found, regardless of the algorithm used or the initial configuration. Although the different methods have different precision and convergence speed, the results were always mutually consistent. Additionally, the independence of the results on the initial configuration implies the results obtained were always a global minimum of the free energy.

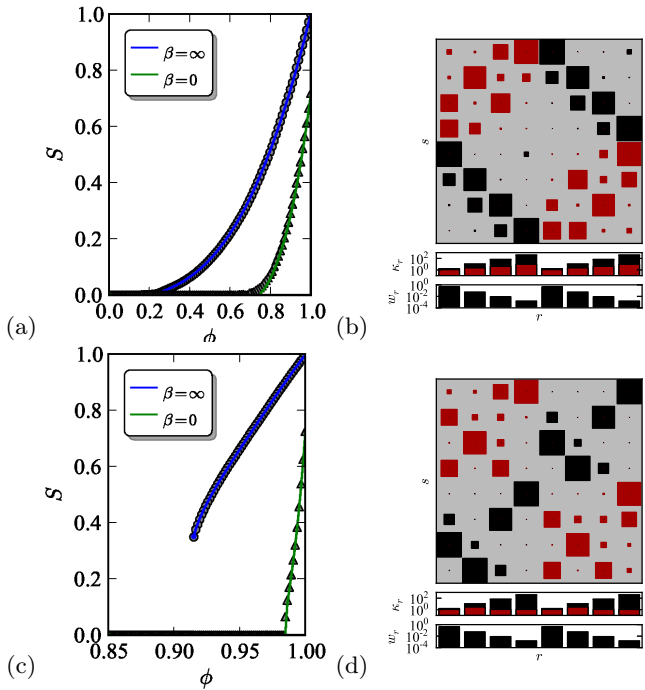

Here we expand briefly on these results by introducing interdependence, i.e. $\langle \hat{k} \rangle \geq 1$. We do so in slightly different way than before, by imposing a separation of the network into two parts of equal size, such that connectivity (interdependence) edges are only allowed between nodes in the same (different) part. This is to mimic, e.g. the situation of computer and electricity infrastructure, where interdependence is only meaningful between the two different types of nodes. With this constraint, we obtain a configuration very similar to the case without interdependence (see Fig. 2), with dissortative and assortative connectivity for the random and targeted removal, respectively. However, the interdependence connectivity is slightly more elaborate: For random failures, the values of $\hat{\kappa}_r$ are correlated with $\kappa_r$, and the $\hat{e}_{rs}$ matrix displays a more assortative connectivity, with blocks depending on other blocks with similar average degree. In the case of targeted attacks, the interdependence shows a CP structure for blocks with low degree, but otherwise follows the same segregated pattern as the connectivity edges.

FIG. 2. (a, c) Fraction $S$ of nodes in the macroscopic component as function of the dilution $\phi$, for different $\beta$ values, fixed values of $w_r$ and $\kappa_r$, and $\langle \hat{k} \rangle = 1.2$, for random failures (a) and targeted attacks (c). The solid lines are results obtained via Eqs. 2 and 4, and the filled symbols are obtained empirically with a network realization of size $N = 10^6$; (b, d) Correlation matrices $e_{rs}/e_r e_s$ (black) and $\hat{e}_{rs}/\hat{e}_r\hat{e}_s$ (red), block size $w_r$ and average degrees $\kappa_r$ (black) and $\hat{\kappa}_r$ (red), for $\beta \to \infty$, for random failures (b) and targeted attacks (d).

## IMPOSING DEGREE CONSTRAINTS

As mentioned in the main text, in some realistic situations there are degree constraints restricting the accessible topologies [9]. Here we introduce it by forcing that the blocks have prescribed sizes and average degrees, following $w_r \propto \kappa_r^{-\gamma}$ and $\kappa_r = e^{cr}$, with $r \in [0, B-1]$ and $c = \ln \kappa_{\max}/(B-1)$. This allows us to have a very broad range of degrees, similar to what is found in scale-free networks. In Fig. 5 of the main text are shown the results for $B = 6$, $\gamma = 2$ and $\kappa_{\max} = 10^3$. Very similar results are found by varying $B$, $\gamma$ and $\kappa_{\max}$ (not shown).

* tiago@itp.uni-bremen.de
† bornholdt@itp.uni-bremen.de


\end{widetext}

\end{document}